\documentclass[twocolumn]{aastex62}
\received{}
\revised{}
\accepted{}
\submitjournal{ApJ Letter}

\shorttitle{Signatures of a Distant Planet around CI~Tau}
\shortauthors{Konishi et~al.}

\begin{document}

\title{Probing Signatures of a Distant Planet around the Young T-Tauri Star CI~Tau Hosting a Possible Hot Jupiter}

\correspondingauthor{Mihoko Konishi}
\email{mihoko.konishi@nao.ac.jp}

\author{Mihoko Konishi}
\author{Jun Hashimoto}
\author[0000-0003-4676-0251]{Yasunori Hori}
\affiliation{Astrobiology Center, National Institute of Natural Science}
\affiliation{National Astronomical Observatory of Japan}



\begin{abstract}
We search for signatures of a distant planet around the two-million-year-old classical T-Tauri star CI~Tau hosting a hot Jupiter candidate ($M_{\rm p}\sin{i} \sim 8.1~M_{\rm Jupiter}$) in an eccentric orbit ($e \sim$0.3). To probe the existence of an outer perturber, we reanalyzed 1.3~mm dust continuum observations of the protoplanetary disk around CI~Tau obtained by the Atacama Large Millimeter/submillimeter Array. We found a gap structure at $\sim$0\farcs8 in CI~Tau's disk. Our visibility fitting assuming an axisymmetric surface brightness profile suggested that the gap is located at a deprojected radius of 104.5$\pm$1.6~au and has a width of 36.9$\pm$2.9~au. The brightness temperature around the gap was calculated to be $\sim$2.3~K lower than that of the ambient disk. Gap-opening mechanisms such as secular gravitational instability and dust trapping can explain the gap morphology in the CI~Tau~disk. The scenario that an unseen planet created the observed gap structure cannot be ruled out, although the coexistence of an eccentric hot Jupiter and a distant planet around the young CI~Tau would be challenging for gravitational scattering scenarios. The mass of the planet was estimated to be between $\sim$0.25~$M_{\rm Jupiter}$ and $\sim$0.8~$M_{\rm Jupiter}$ from the gap width and depth (0.41$^{+0.04}_{-0.06}$) in the modeled surface brightness image, which is lower than the current detection limits of high-contrast direct imaging. The young classical T-Tauri CI~Tau may be a unique system to explore the existence of a potential distant planet as well as the origin of an eccentric hot Jupiter. 
\end{abstract}

\keywords{
planets and satellites: formation, 
stars: individual (CI~Tau), 
protoplanetary disks, 
planet-disk interactions,  
submillimeter: stars 
}


\section{Introduction}\label{sec:intro}
Planets are forming in the circumstellar environment around young stars with an age of $\lesssim$10 million years (Myrs). The detection of planets orbiting a young star can be decisive evidence for the link between planet formation and orbital evolution. However, the fast rotation and high stellar activity on the surface of a young star inhibit unveiling the presence of planets. Despite such difficulties in planet hunting around young stars, four hot Jupiters orbiting around T-Tauri stars have been discovered recently: CVSO~30 which is in dispute \citep{vaneyken12}, V830~Tau~b \citep{donati16}, TAP~26~b \citep{yu17}, and CI~Tau~b \citep{john16}. Close-in gas giants that formed further out likely experienced Type II migration because of disk--planet interactions, or were perturbed into the proximity of their star by secular perturbations of other planets or stellar companions, namely, planet--planet scattering \citep[e.g.,][]{rasio96} and the Kozai--Lidov mechanism \citep{kozai62}. Gravitational perturbations from others cause excitation of the eccentricity and inclination of a scattered planet. A young hot Jupiter that remains in an eccentric orbit should keep the formation record because the time scales of orbital decay and circularization resulting from tidal forces exerted on a planet are of the order of $\sim$Gyr.

CI~Tau is a 2-Myr-old classical T-Tauri star \citep{herc14} in the Taurus star-forming region (e.g., \citealp{kraus+2007}) with a heliocentric distance of $\sim$140~pc (e.g., \citealp{wichman+1998,torres+2012}). We employ recent astrometic measurements from the $Gaia$ data release 2 which revealed the distance of 158.7$\pm$1.2~pc \citep{bailer-jones+18}. Its mass was estimated as 0.8~$M_{\odot}$ \citep{guil14}. \citet{john16} reported an eccentric hot Jupiter having an orbital period of $\sim$9~days and $M_{\rm planet}\sin{i} = 8.1~M_{\rm Jupiter}$ around CI~Tau. Recently, $K2$ photometry has revealed two periodic variations of $\sim$6.6 and $\sim$9~days in the lightcurves of CI~Tau \citep{bidd18}. According to the disk geometry ($i = 45.7^\circ$) of CI~Tau derived from 1.3~mm continuum observations \citep{guil14}, CI~Tau~b is expected to be a non-transiting planet and the stellar rotation period corresponds to periodic signals of $6.6$~days. The observed 9-day photometric variability can be interpreted as the periodic dimming of light by planet--disk interactions such as pulsed accretion \citep{bidd18}. In fact, mass accretion from a disk onto CI~Tau measured $\sim 3 \times 10^{-8}~M_{\odot}~{\rm yr}^{-1}$ with HI Pfund $\beta$ emission \citep{saly13} and Brackett $\gamma$ lines \citep{mcclure13} as a tracer of accretion flow. 

The presence of an outer companion can be directly proved by high-contrast imaging. Previous binary surveys revealed no stellar companions around CI~Tau (e.g., \citealp{kraus+2007}). High-contrast imaging searching for distant gas giants also has failed to detect faint point sources within their detection limit of $\sim$18~mag at 1.6~$\mu$m \citep{uyam17}. The presence of companions could also be suggested indirectly from the morphology of the circumstellar disk, if companions embedded in the disk can generate a gap structure and spiral arms in the disk. However, previous observations of the submillimeter continuum revealed no gap structure in the disk of CI~Tau \citep[e.g.,][]{andr07}. In addition, there is no significant deficit of infrared (IR) emission in spectral energy distributions (SED; see \citealt{andr07} and references therein), indicating that CI~Tau does not exhibit properties of so-called transitional disks \citep{espa14}. Hence, such observational features of CI~Tau invoke a different mechanism for the eccentricity pumping of CI~Tau~b, instead of planet--planet scattering. The high eccentricity of CI~Tau~b can be mediated by disk-driven eccentricity growth \citep{rosotti17}. \citet{ragusa18} showed that the eccentricity of a CI~Tau~b-like planet grows linearly to be $\sim 0.1$ while oscillating periodically, although it may continue to increase. Recently, the Atacama Large Millimeter/submillimeter Array (ALMA) has successfully revealed spatially resolved multiple-ring structures in the disks of HL~Tau \citep{alma15} and HD~163296 \citep{isel16}. Like CI~Tau, HL~Tau and HD~163296 have not been identified as multi-ring disks until ALMA discoveries because these two have no IR deficit in their SED. Therefore, we revisit the disk structure of CI~Tau and probe signatures of a distant planet around CI~Tau by utilizing ALMA high-resolution observations, which are publicly available from the ALMA data archive. 

In this paper, we present the ALMA observations and results of our data reduction in Section~\ref{sec:obsres}. The disk model of the CI~Tau disk and results of our visibility fitting are shown in Section~\ref{sec:model}. We discuss possible gap-opening mechanisms of the CI~Tau system in Section~\ref{sec:discuss}. Section~\ref{sec:sum} is a brief summary of this study.

\section{Observations and data reduction}\label{sec:obsres}
\subsection{Observations}
CI~Tau was observed by ALMA band 6 (1.3~mm) as part of the Cycle 3 program 2015.1.01207.S (Principal Investigator: H. Nomura) on August 27, 2016 UT. During the observation, 44 antennas in the 12-m array were used with a baseline length ranging from 15.1~m to 1.6~km. The ALMA array correlator was configured with three spectral windows (937.500~MHz bandwidth) centered on 216.7, 231.2, 234.4~GHz and the other one (234.375~MHz bandwidth) centered on 219.9~GHz to cover H$_2$S, $^{13}$CS(5$-$4), SO$_2$(16$_{6,\,10}-$17$_{5,\,13}$), and SO(6$_5-$5$_4$) lines. The four spectral windows have velocity resolutions of 0.675, 0.633, 1.249, and 0.666~km~s$^{-1}$, respectively. The precipitable water vapor was $\sim$1.9~mm during the observation. The quasar J0510$+$1800 was observed for bandpass and flux calibration, and J0431$+$2037 was used for phase calibration. The total on-source integration time was 18.4~min. 

\subsection{Data Reduction}
Data calibration was performed using the Common Astronomy Software Applications (CASA) pipeline version 4.7.0 \citep{mcmu07} according to the steps shown in the calibration scripts provided by ALMA. We used only phase self-calibration because the signal-to-noise (S/N) ratio remained nearly unchanged by the amplitude calibration. 

\begin{figure*}
    \centering
    \includegraphics[clip,width=\linewidth]{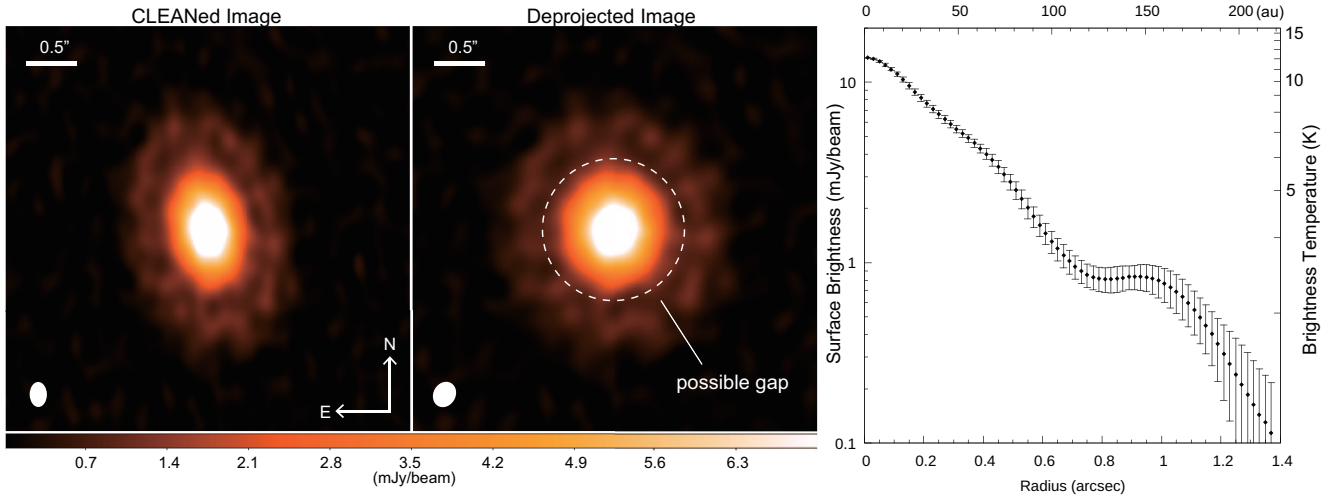}
    \caption{ALMA 1.3~mm dust continuum image of the CI~Tau disk. (Left panel) A CLEANed image with superuniform weighting. The beam size is 0\farcs25$\times$0\farcs17 with a position angle of 0.6$^{\circ}$, as shown in white. (Middle panel) Deprojected image obtained by using $i=44.6^{\circ}$ and $PA=11.8^{\circ}$. The beam size is 0\farcs26$\times$0\farcs22 with a position angle of 150.3$^{\circ}$. The dashed line indicates a possible gap structure at $\sim$0\farcs8. (Right panel) Deprojected radial profile of surface brightness averaged azimuthally. The heliocentric distance of 158.7$\pm$1.2~pc was used to convert arcseconds to au. 
    }\label{fig_finimg}
\end{figure*}

The dust continuum image (with a total bandwidth of 1875.0~MHz) was generated from line-free spectral windows by using the \verb#CLEAN# algorithm \citep{rau11}. The left panel of Figure~\ref{fig_finimg} shows the \verb#CLEAN#ed image with ``superuniform'' weighting. The beaming size of the \verb#CLEAN#ed image is 0\farcs25~$\times$~0\farcs17 (40~$\times$~27~au) with a position angle of 0.6$^{\circ}$, and its 1$\sigma$ noise is $\sim$0.17~mJy~beam$^{-1}$. To minimize the beam-elongation effect, we generated the \verb#CLEAN#ed image with a circular beam size of 0\farcs29~$\times$~0\farcs28 using \verb#uvtaper# in the CASA tools. We fitted the 20$\sigma$-contour level of the \verb#CLEAN#ed image with an elliptical shape using orthogonal distance regression, obtaining values of $i=$44.6$^{\circ}$ and $PA=$11.8$^{\circ}$, where $i$ and $PA$ are the system inclination and the position angle, respectively. The middle panel of Figure~\ref{fig_finimg} shows the \verb#CLEAN#ed image with the observed visibilities deprojected in the $uv$ plane, using the following equations \citep[e.g.,][]{zhan16}: 
 $u'    =   (u \cos{PA} - v \sin{PA}) \times \cos{i}$ and
 $v'    =   (u \sin{PA} - v \cos{PA})$, 
where we used the measured $i$ and $PA$. Finally, we calculated the surface brightness profile by averaging the deprojected image azimuthally. The right panel of Figure~\ref{fig_finimg} shows the surface brightness as a function of deprojected radius, and the brightness temperature was calculated using the Planck function. We found a small dip at $\sim$120~au, indicating a possible gap structure in the CI~Tau disk. 

To examine the morphology of the gap structure, we identified positions of a ``valley'' on the deprojected image every five degrees of azimuthal angle by fitting an inverse Gaussian function to the surface brightness profile. We determined an elliptical shape by using the valley's positions following the same method as that in the previous paragraph. Then, we found the eccentricity of a possible gap to be $\sim$0.46. However, the low S/N ratio of 1.3~mm dust continuum observations may overestimate the gap eccentricity. Consequently, we consider the gap as an axisymmetric circular shape in this paper.

\section{Disk modeling}\label{sec:model}
The elongated beam is slightly larger in size than the gap, which inhibits determining the gap width and depth (see Figure~\ref{fig_finimg}). To derive the geometric structure of the gap from the overall surface brightness profile, we considered a radial profile of the surface brightness $I_\nu (R)$ that can reproduce the observed disk structure. As seen in Figure~\ref{fig_finimg}, the brightness temperature falls below the dust temperature at the disk midplane given in \citet{andr07}, $T_d(R) = 178\pm3\ (R/1\ {\rm au})^{-0.57\pm0.02}$, where $R$ is the radial distance from the star. This indicates that thermal emissions from dust grains in the CI~Tau disk are optically thin and the surface brightness can be approximately expressed as 
 $I_{\nu} \propto B_{\nu}(T_{d})(1-e^{-\tau}) \propto T_{d}\,\tau,$
where $B_{\nu}$ is the blackbody intensity and $\tau$ is the optical depth. The optical depth can be defined as $\tau = \kappa\Sigma$, where $\kappa$ and $\Sigma$ denote the opacity and the surface density of dust, respectively. We used an exponentially tapered  power-law distribution of the surface dust density predicted from viscous accretion disks \citep{lynd74,hart98}, obtaining the following relation: 
\begin{eqnarray*}
  I_\nu (R) \propto \left(\frac{R}{R_{c}}\right)^{-(q+\gamma)}{\rm exp}\left[-\left(\frac{R}{R_{c}}\right)^{2-\gamma}\right], 
\end{eqnarray*}
where $R_{\rm c}$ is a characteristic radius, $\gamma$ and $q$ specify the radial dependence of the disk alpha viscosity $\nu_\alpha \propto R^\gamma$ and the dust temperature $T_{\rm d} \propto R^{-q}$. In this study, we adopted $q = 1/2$, as suggested by \citet{andr07}, and the inclination ($i=$44.6$^{\circ}$) and the position angle ($PA=$11.8$^{\circ}$) derived from the \verb#CLEAN#ed image.

We introduced three parameters to describe the gap structure: the inner edge ($R_{\rm gap\_in}$), the outer edge ($R_{\rm gap\_out}$), and the depth of the gap ($\delta$). To model the surface brightness profile of the observed CI~Tau disk with a gap, we defined six parameters in total, as stated above ($R_{\rm gap\_in}$, $R_{\rm gap\_out}$, $\gamma$, $\delta$, $R_{c}$, and $F_{\rm total}$), where we used the integral of surface brightness with respect to solid angle, i.e., the total flux density ($F_{\rm total}$). We generated the modeled image from the surface brightness profile that compensated for distortion from the inclination. The modeled image was converted to complex visibilities using the public Python code {\sf vis\_sample}\footnote{{\sf vis\_sample} is publicly available at {\sf https://github.com/AstroChem/vis\_sample} or from the Anaconda Cloud at {\sf https://anaconda.org/rloomis/vis\_sample}}. We obtained the azimuthally averaged profile of the complex visibilities, which were deprojected in the $uv$ plane \citep[e.g.,][]{zhan16}.

\begin{figure*}
    \centering
    \includegraphics[clip,width=\linewidth]{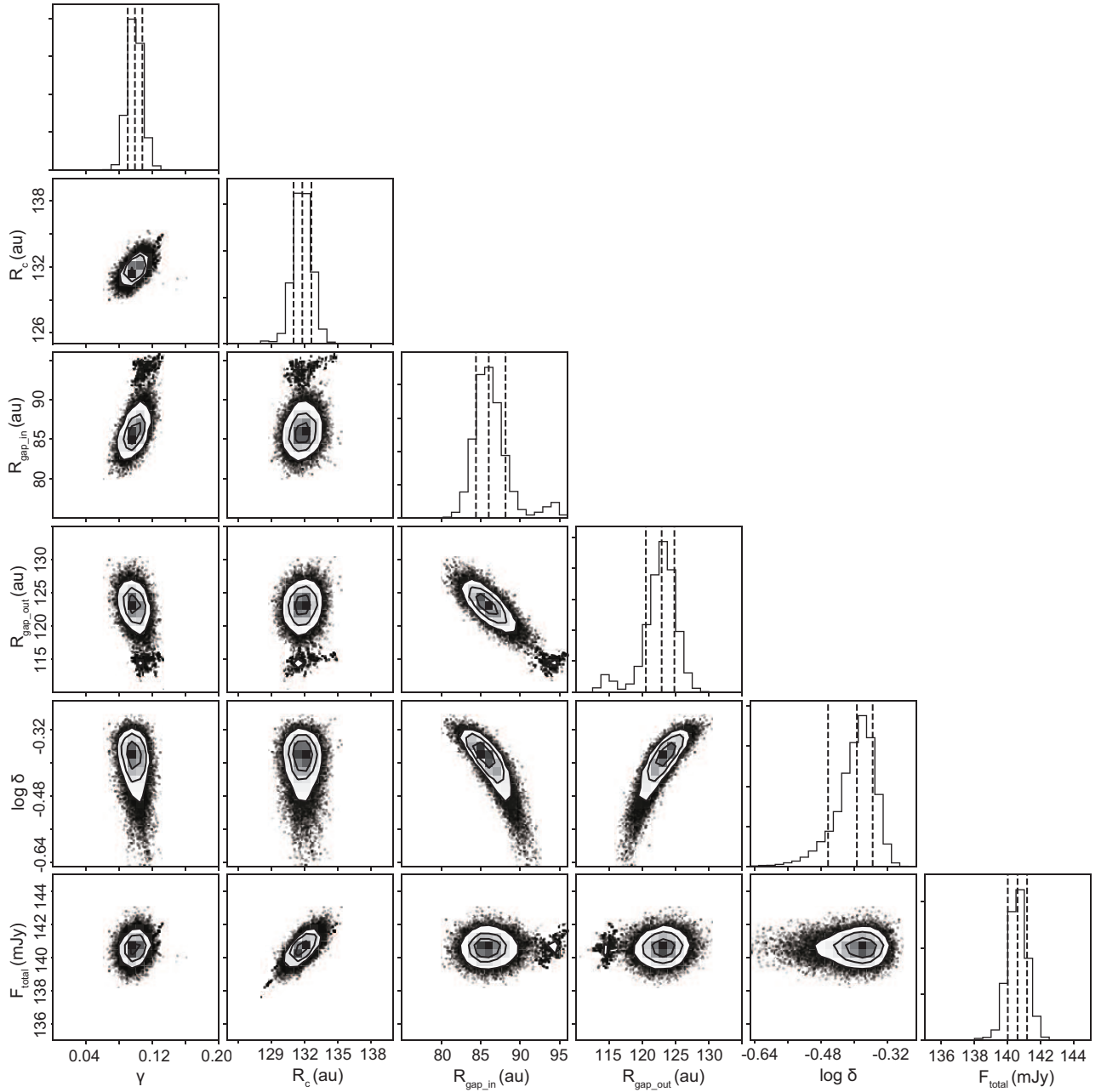}
    \caption{Corner map of our MCMC results. The dashed lines indicate $-$1$\sigma$, most likely, and 1$\sigma$ points from the left to right. There are no local concentrated regions. 
    }\label{fig_mcmcres}
\end{figure*}

\begin{figure*}
    \centering
    \includegraphics[clip,width=\linewidth]{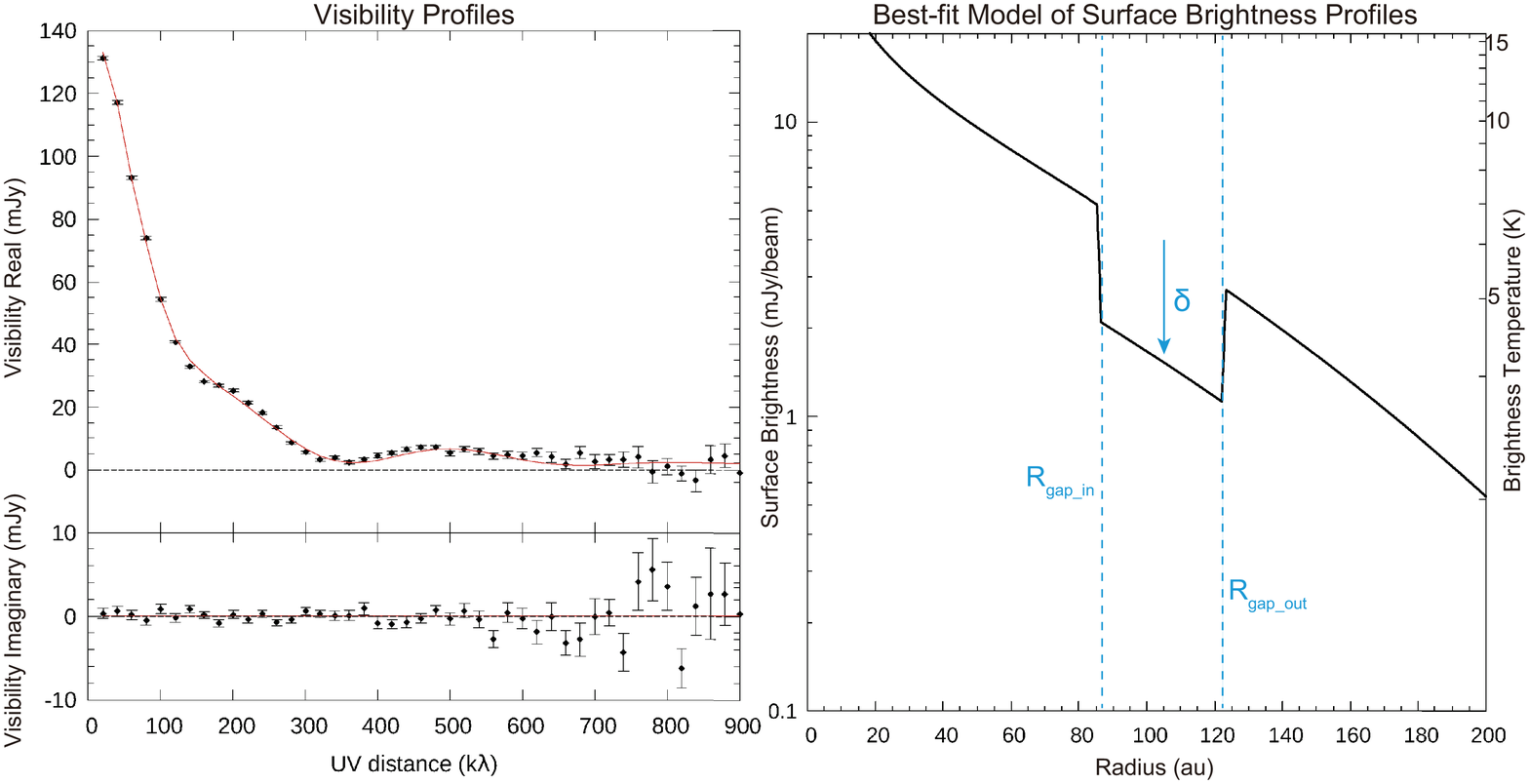}
    \caption{(Left panel) Azimuthally averaged visibility profile after deprojection in the $uv$ plane. The top and bottom panels are the real and imaginary parts of the CI~Tau visibilities, respectively. The black diamonds represent the observed visibilities with 1$\sigma$ error bars. The red lines indicate the best-fit ones found in Section~\ref{sec:model}. (Right panel) Surface brightness profile obtained using our best-fit values. Note that the measured surface brightness profile shown in Figure~\ref{fig_finimg} was deblured by effects of the finite beam size.
    }\label{fig_vis}
\end{figure*}

Then, we used the Markov Chain Monte Carlo (MCMC) method to fit our disk model to the radial profile of observed complex visibilities, using the {\sf emcee} package \citep{foreman-mackey+2013}. We used flat prior distributions for each parameter. The best-fit values with the reduced $\chi^{2}$ values of $\sim$4.7 and parameter ranges for the MCMC fitting are summarized in Table~\ref{tab:table1}. In the MCMC analysis, we excluded 500 steps as the burn-in phase and then, ran 1000 iterations with 100 walkers for convergence. The 1000 trials for the MCMC fitting are plotted in Figure~\ref{fig_mcmcres}. Figure~\ref{fig_vis} shows the visibility and surface brightness profiles with the best-fit parameters. We confirmed that there are no locally concentrated regions for our MCMC sampling, as shown in the corner map (see Figure~\ref{fig_mcmcres}). We found that the gap is located at 104.5$\pm$1.6~au (0\farcs66$\pm$0\farcs01) and it has a width of 36.9$\pm$2.9~au and a depth of 0.41$^{+0.04}_{-0.06}$. 

\begin{deluxetable*}{cccccc}
\tablecaption{Results of our MCMC fitting \label{tab:table1}}
\tablehead{
\colhead{$\gamma$} & \colhead{$R_{\rm gap\_in}$} & \colhead{$R_{\rm gap\_out}$} & \colhead{$R_{\rm c}$} & \colhead{log $\delta$} & \colhead{$F_{\rm total}$} \\
\colhead{}         & \colhead{(au)}              & \colhead{(au)}              & \colhead{(au)}         & \colhead{}             & \colhead{(mJy)} 
}
\startdata
0.10$^{+0.01}_{-0.01}$  & 86.0$^{+2.2}_{-1.7}$ & 122.9$^{+2.0}_{-2.5}$ & 131.8$\pm${+1.0} & $-$0.39$^{+0.04}_{-0.07}$ & 140.6$^{+0.6}_{-0.6}$ \\
$[$0.0, 2.0$]$            & $[$31.7, 127.0$]$      & $[$63.5, 158.7$]$      & $[$47.6, 142.8$]$       & $[-$3.0, 0.0$]$ & $[$100.0, 200.0$]$ \\
\enddata
\tablecomments{
 Parentheses describe parameter ranges in our MCMC calculations.}
\end{deluxetable*}

\section{Discussion}\label{sec:discuss}

We have presented ALMA 1.3~mm continuum observations of a protoplanetary disk around CI~Tau. The dust continuum map shown in Figure \ref{fig_finimg} reveals a ring-shaped structure with a width of $\sim$37~au located at a radius of $\sim$104~au. There are various ideas that can produce an axisymmetric dust gap in a circumstellar disk. First, we consider dust growth and pile-up near the sublimation front of volatile molecules resulting from ice condensation and sintering effects \citep[e.g.,][]{okuzumi+16}. Among the major volatile species (H$_2$O, CO, CO$_2$, CH$_4$, H$_2$S, N$_2$, CH$_4$, and CH$_3$OH), CO, CH$_4$, and N$_2$ condense into ice at extremely low temperature in a protoplanetary disk, at $\sim$20--30~K \citep[e.g.,][]{2014A&A...570A..35M}. According to the estimations from submillimeter continuum observations combined with the SED \citep{andr07}, the radial temperature profile in CI~Tau's disk follows $T(R) = 178\pm3\,(R/1~{\rm au})^{-0.57\pm0.02}$, indicating that the gas temperature near the gap is $\sim$12.6~K. This implies that a ring-like structure lies beyond the snow lines of the three molecules. Thus, the growth and concentration of dust aggregates near the snow line could not explain formation of the observed gap in the disk around CI~Tau.

A passive disk such as CI~Tau ($\dot{M} \sim 10^{-8}~M_\odot~{\rm yr^{-1}}$) becomes thermally unstable against stellar irradiation (e.g., \citealp{watanabe08}). Thermal waves that are excited in an outer region of a flared disk produce fluctuations of temperature. \citet{watanabe08} found that temperature differences at the disk midplane between the illuminated and shadow regions are $\sim$10~K at $\sim$80~au. Our best-fit model of surface brightness profiles indicates that the brightness temperature near the gap is $\sim$2.3~K lower than that of the ambient disk (see the right panel of Figure~\ref{fig_vis}). Since thermal waves propagate inward and their amplitudes damp out, fluctuations of midplane temperature can weaken in the inner disk. Thermal waves in an irradiated disk are, however, incompatible with the small temperature gradient in the outer region, such as the location of the gap seen in CI~Tau's disk.

The disk gas in a cold region can be ionized by cosmic rays and stellar X-ray irradiation, which may drive the magneto-rotational instability (MRI; e.g., \citealp{balbus91}). Local pressure bumps that occur in a MRI-active region or non-ideal magnetohydrodynamics (MHD) effects such as the Hall effects and ambipolar diffusion which suppress turbulence can trap dust particles, leading to a concentration of dust particles. Dust settling can also gather dust into a toroidal vortex produced by a baroclinic instability \citep{loren-aguilar+15}. Consequently, a hydrodynamic instability resulting from interactions between disk gas and dust particles can be a possible mechanism for generating the ring-like structure seen in the CI~Tau disk.

Previous results of CI~Tau observations indicated weak turbulence predicted from the small accretion rate (e.g., \citealp{saly13}) and a high dust-to-gas mass ratio \citep[$\sim$10:][]{williams+14} in a gravitationally stable disk. Disk properties are favorable for triggering secular gravitational instability (secular GI) at a radius of $\gtrsim$80~au. \citet{takahashi14} predicted that the most unstable wavelength is dozens of au at $\sim$80--100~au for a weakly turbulent disk ($\alpha \sim 10^{-4}$), which is consistent with the observed gap structures. The secular GI as well as the MRI can accumulate dust particles and promote formation of a dusty ring in the disk of CI~Tau. 

Finally, we discuss the possibility of a planet embedded in a disk. A massive planet can clear out the surrounding disk and produce a ring structure seen in CI~Tau's disk. Using the estimated aspect ratio of 0.122 and empirical formulas for gap-opening induced by a planet \citep{kanagawa+2015,kanagawa+2016}, we find the mass of a planet to be 0.8~$M_{\rm Jupiter}$ and 0.25~$M_{\rm Jupiter}$, where we consider $\alpha = 10^{-3}$ and $10^{-4}$ as the strength of turbulent viscosity in the CI~Tau disk, respectively. It is noted that we should take some cautions to interpret the derived planet mass when such gas surface density formulas are applied to dust surface density. The origin of a distant planet is
(i) a small core scattered from an inner region accreting the disk gas in situ \citep{piso+14,kikuchi+14} 
and (ii) a by-product of gravitational scattering among planets. 
Assuming that CI~Tau~b formed at 5--10~au, the Lidov--Kozai oscillation can be initiated by an outer companion with a mass of 0.25--0.8~$M_{\rm Jupiter}$ after 40--380~Myr \citep[e.g.][]{holman+94}. Planet-planet scattering also likely occurs in gas-free environment. The coexistence of the distant planet and the eccentric hot Jupiter around CI~Tau would be challenging for gravitational scattering scenarios. Previous high-contrast imaging observations found no point-like sources with $\lesssim$18~mag in the $H$ band at 0\farcs75 \citep{uyam17}, which corresponds to a planet with a mass of $\gtrsim$3~$M_{\rm Jupiter}$ around CI~Tau aged $2$~Myr based on the COND models \citep{baraffe+03}, because the predicted mass of the planet creating the gap is lower than their current detection limits. Therefore, further deep high-contrast imaging using the extreme Adaptive Optics in near-infrared wavelengths (e.g., Subaru/CHARIS, Gemini/GPI, and VLT/SPHERE) and/or ALMA in (sub-)millimeter wavelengths will be able to verify the existence of the planet at the gap and to constrain signatures of other planets. 

\section{Summary}\label{sec:sum}

We searched for signatures of an outer planetary companion to CI~Tau harboring a possible hot Jupiter CI~Tau~b by using the ALMA archive data. The 1.3~mm dust continuum revealed a gap structure at $\sim$0\farcs8. Our visibility fitting found that the gap located at 104.5$\pm$1.6~au has a width of 36.9$\pm$2.9~au and a depth of 0.41$^{+0.04}_{-0.06}$. Dust trapping and the secular GI can explain the ring-like structures in the CI~Tau disk. The existence of an outer planet can also explain the observed gap structure, although the coexistence of an eccentric hot Jupiter and a distant planet around the young CI~Tau would be challenging for gravitational scattering scenarios. The planet's mass estimated from the dust gap structure is between $\sim$0.25~$M_{\rm Jupiter}$ and $\sim$0.8~$M_{\rm Jupiter}$, which is consistent with non-detection of point-like sources with a mass of $\gtrsim$3~$M_{\rm Jupiter}$ by previous observations. Further high-contrast imaging will verify the existence of the planet embedded in the gap and constrain signatures of other planets in the CI~Tau disk. The young classical T-Tauri CI~Tau may be a unique system to explore the existence of a potential distant planet as well as the origin of an eccentric hot Jupiter.

\acknowledgments
We would like to express thanks to Hideko Nomura and Aya Higuchi for their efforts to obtain the ALMA data. 
This paper makes use of the following ALMA data: ADS/JAO.ALMA\#2015.1.01207.S. ALMA is a partnership of ESO (representing its member states), NSF (USA), and NINS (Japan), together with NRC (Canada) and NSC and ASIAA (Taiwan) and KASI (Republic of Korea), in cooperation with the Republic of Chile. The Joint ALMA Observatory is operated by ESO, AUI/NRAO and NAOJ. 
We also thank the anonymous referee for giving helpful comments to improve our manuscript. J.H acknowledges support from JSPS KAKENHI Grant Number 17K14258.

%
\vspace{5mm}
\facilities{ALMA}

\software{CASA (https://casa.nrao.edu/),  
          {\sf vis\_sample} (https://github.com/AstroChem/vis\_sample or https://anaconda.org/rloomis/vis\_sample), 
          {\sf emcee} \citep{foreman-mackey+2013}
          }


\begin{thebibliography}{}
\bibitem[ALMA Partnership et~al.(2015)]{alma15} ALMA Partnership, Brogan, C.~L., P\'erez, L.~M., et~al. 2015, \apjl, 808, L3
\bibitem[Andrews \& Williams(2007)]{andr07} Andrews, S.~M., \& Williams, J.~P. 2007, \apj, 659, 705 
\bibitem[Bailer-Jones et~al.(2018)]{bailer-jones+18} Bailer-Jones, C.~A.~L., Rybizki, J., Fouesneau M., Mantelet, G. \& Andrae, R. 2018, arXiv:1804.10121 
\bibitem[Balbus \& Hawley(1991)]{balbus91} Balbus, S.~A., \& Hawley, J.~F. 1991, \apj, 376, 214
\bibitem[Baraffe et~al.(2003)]{baraffe+03} Baraffe, I., Chabrier, G., Barman, T.~S., Allard, F., \& Hauschildt, P.~H. 2003, \aap, 402, 701
\bibitem[Biddle et~al.(2018)]{bidd18} Biddle, L.~I., Johns-Krull, C.~M., Llama, J., Prato, L., \& Skiff, B.~A. 2018, \apjl, 853, L34
\bibitem[Donati et~al.(2016)]{donati16} Donati, J.~F., Moutou, C., Malo, L., et~al. 2016, \nat, 534, 662
\bibitem[Espaillat et~al.(2014)]{espa14} Espaillat, C., Muzerolle, J., Najita, J., et~al. 2014, Protostars and Planets VI, 497
\bibitem[Foreman-Mackey et~al.(2013)]{foreman-mackey+2013} Foreman-Mackey, D., Hogg, D.~W., Lang, D., \& Goodman, J. 2013, \pasp, 125, 306
\bibitem[Guilloteau et~al.(2014)]{guil14} Guilloteau, S., Simon, M., Pi\'etu, V., et~al. 2014, \aap, 567, A117
\bibitem[Hartmann et~al.(1998)]{hart98} Hartmann, L., Calvet, N., Gullbring, E., \& D'Alessio, P. 1998, \apj, 495, 385
\bibitem[Herczeg \& Hillenbrand(2014)]{herc14} Herczeg, G.~J., \& Hillenbrand, L.~A. 2014, \apj, 786, 97
\bibitem[Holman et al.(1997)]{holman+94} Holman, M., Touma, J., \& Tremaine, S.\ 1997, \nat, 386, 254 
\bibitem[Isella et~al.(2016)]{isel16} Isella, A., Guidi, G., Testi, L., et~al. 2016, Phys. Rev. Lett., 117, 251101
\bibitem[Johns-Krull et~al.(2016)]{john16} Johns-Krull, C.~M., McLane, J.~N., Prato, L., et~al. 2016, \apj, 826, 206
\bibitem[Kanagawa et~al.(2016)]{kanagawa+2016} Kanagawa, K.~D., Muto, T., Tanaka, H., et~al. 2016, \pasj, 68, 43
\bibitem[Kanagawa et~al.(2015)]{kanagawa+2015} Kanagawa, K.~D., Tanaka, H., Muto, T., Tanigawa, T., \& Takeuchi, T. 2015, \mnras, 448, 994
\bibitem[Kikuchi et~al.(2014)]{kikuchi+14} Kikuchi, A., Higuchi, A., \& Ida, S. 2014, \apj, 797, 1
\bibitem[Kozai(1962)]{kozai62} Kozai, Y. 1962, \aj, 67, 579
\bibitem[Kraus \& Hillenbrand(2007)]{kraus+2007} Kraus, A.~L., \& Hillenbrand, L.~A. 2007, \apj, 662, 413
\bibitem[Kretke \& Lin(2007)]{kretke+07} Kretke, K.~A., \& Lin, D.~N.~C. 2007, \apjl, 664, L55
\bibitem[Lor\'en-Aguilar \& Bate(2015)]{loren-aguilar+15} Lor\'en-Aguilar, P., \& Bate, M.~R. 2015, \mnras, 453, L78
\bibitem[Lynden-Bell \& Pringle(1974)]{lynd74} Lynden-Bell, D., \& Pringle, J.~E. 1974, \mnras, 168, 603
\bibitem[Marboeuf et~al.(2014)]{2014A&A...570A..35M} Marboeuf, U., Thiabaud, A., Alibert, Y., Cabral, N., \& Benz, W. 2014, \aap, 570, A35
\bibitem[McClure et~al.(2013)]{mcclure13} McClure, M.~K., Calvet, N., Espaillat, C., et~al. 2013, \apj, 769, 73
\bibitem[McMullin et~al.(2007)]{mcmu07} McMullin, J.~P., Waters, B., Schiebel, D., Young, W., \& Golap, K. 2007, in Astronomical Society of the Pacific Conference Series, Vol. 376, Astronomical Data Analysis Software and Systems XVI, ed. R. A. Shaw, F. Hill, \& D. J. Bell, 127
\bibitem[Okuzumi et~al.(2016)]{okuzumi+16} Okuzumi, S., Momose, M., Sirono, S.-i., Kobayashi, H., \& Tanaka, H. 2016, \apj, 821, 82
\bibitem[Piso \& Youdin(2014)]{piso+14} Piso, A.-M.~A., \& Youdin, A.~N. 2014, \apj, 786, 21
\bibitem[Ragusa et~al.(2018)]{ragusa18} Ragusa, E., Rosotti, G., Teyssandier, J., et~al. 2018, \mnras, 474, 4460
\bibitem[Rasio \& Ford et~al.(1996)]{rasio96} Rasio, F.~A., \& Ford, E. B. 1996, Science, 274, 954
\bibitem[Rau \& Cornwell(2011)]{rau11} Rau, U., \& Cornwell, T.~J. 2011, \aap, 532, A71
\bibitem[Rosotti et~al.(2017)]{rosotti17} Rosotti, G.~P., Booth, R.~A., Clarke, C.~J., et~al. 2017, \mnras, 464, L114
\bibitem[Salyk et~al.(2013)]{saly13} Salyk, C., Herczeg, G.~J., Brown, J.~M., et~al. 2013, \apj, 769, 21
\bibitem[Takahashi \& Inutsuka(2014)]{takahashi14} Takahashi, S.~Z., \& Inutsuka, S.-i. 2014, \apj, 794, 55
\bibitem[Torres et~al.(2012)]{torres+2012} Torres, R.~M., Loinard, L., Mioduszewski, A.~J., et~al. 2012, \apj, 747, 18
\bibitem[Uyama et~al.(2017)]{uyam17} Uyama, T., Hashimoto, J., Kuzuhara, M., et~al. 2017, \aj, 153, 106
\bibitem[van Eyken et~al.(2012)]{vaneyken12} van Eyken, J.~C., Ciardi, D.~R., von Braun, K., et~al. 2012, \apj, 755, 42
\bibitem[Watanabe \& Lin(2008)]{watanabe08} Watanabe, S.-i., \& Lin, D.~N.~C. 2008, \apj, 672, 1183
\bibitem[Wichmann et~al.(1998)]{wichman+1998} Wichmann, R., Bastian, U., Krautter, J., Jankovics, I., \& Rucinski, S.~M. 1998, \mnras, 301, L39
\bibitem[Williams \& Best(2014)]{williams+14} Williams, J.~P., \& Best, W.~M.~J. 2014, \apj, 788, 59
\bibitem[Yu et~al.(2017)]{yu17} Yu, L., Donati, J.-F., H\'ebrard, E.~M., et~al. 2017, \mnras, 467, 1342
\bibitem[Zhang et~al.(2016)]{zhan16} Zhang, K., Bergin, E.~A., Blake, G.~A., et~al. 2016, \apjl, 818, L16
\end{thebibliography}
\end{document}